\documentclass[12pt,preprint]{aastex}

\newcommand\msun{\hbox{\,M$_\odot$}}

\newcommand\viss{${\mathcal{V}^2}$ }

\shorttitle{First Visual Orbit for WR~140}
\shortauthors{Monnier et al.}

\begin{document}


\title{First visual orbit for the prototypical colliding-wind binary WR~140}


\author{J.~D.~Monnier\altaffilmark{1}, 
Ming Zhao\altaffilmark{1}, E. Pedretti\altaffilmark{1,9}, R. Millan-Gabet\altaffilmark{2},
J.-P. Berger\altaffilmark{3}, W. Traub\altaffilmark{4}, F.P.Schloerb\altaffilmark{5},
T. ten Brummelaar\altaffilmark{6}, H. McAlister\altaffilmark{6}, S. Ridgway\altaffilmark{7}, 
L. Sturmann\altaffilmark{6}, J. Sturmann\altaffilmark{6}, N. Turner\altaffilmark{6},
F. Baron\altaffilmark{1}, S. Kraus\altaffilmark{1}, A. Tannirkulam\altaffilmark{1,10},  P. M. Williams\altaffilmark{8}
}

\altaffiltext{1}{monnier@umich.edu: University of Michigan Astronomy Department, 941 Dennison Bldg, Ann Arbor, MI 48109-1090, USA.}
\altaffiltext{2}{California Institute of Technology, NASA Exoplanet Science
 Institute, Pasadena, CA 91125, USA}
\altaffiltext{3}{IPAG,CNRs/UMR 5571, Universite J. Fourier, BP-53, F-38041 Grenoble Cedex, France}
\altaffiltext{4}{Jet Propulsion Laboratory, California Institute of Technology, 
M/S 301-355, 4800 Oak Grove Drive, Pasadena, CA 91109, USA}
\altaffiltext{5}{University of Massachusetts, Department of Astronomy, Amherst, MA 01003-4610, USA}
\altaffiltext{6}{The CHARA Array of Georgia State University, Mt. Wilson, CA, 91023, USA}
\altaffiltext{7}{National Optical Astronomy Observatory, 950 N. Cherry Ave, Tucson AZ 85719. USA}
\altaffiltext{8}{Institute for Astronomy, University of Edinburgh, Royal 
Observatory, Blackford Hill, Edinburgh, UK}
\altaffiltext{9}{European Organisation for Astronomical Research in the Southern Hemisphere, Karl Schwarzschild Strasse 2, 85748 Garching bei München}
\altaffiltext{10}{Center for Micro Finance, The Institute for Financial Management and Research}
\email{JDM: monnier@umich.edu}


\begin{abstract}
Wolf-Rayet stars represent one of the final stages of massive stellar
evolution.  Relatively little is known about this short-lived phase
and we currently lack reliable mass, distance, and binarity
determinations for a representative sample.  Here we report the first
visual orbit for WR~140($=$HD193793), a  WC7$+$O5 binary
system known for its periodic dust production episodes triggered by
intense colliding winds near periastron passage. The IOTA and CHARA
interferometers resolved the pair of stars in each year from
2003--2009, covering most of the highly-eccentric, 7.9~year orbit.
Combining our results with the recent improved double-line
spectroscopic orbit of Fahed et al. (2011), we find the WR~140 system
is located at a distance of $1.67\pm0.03$~kpc, composed of a WR star
with $M_{\rm WR} = 14.9\pm0.5\msun$ and an O star with $M_{\rm O}=
35.9\pm1.3\msun$.  Our precision orbit yields key parameters with
uncertainties $\sim\times$6 smaller than previous work and paves the
way for detailed modeling of the system. Our newly measured
flux ratios at the near-infrared H and Ks bands allow an SED
decomposition and analysis of the component evolutionary states.

\end{abstract}





\section{Introduction}
Wolf-Rayet (WR) stars are evolved massive stars characterized by
intense mass-loss through radiation-driven winds.  These hot,
emission-line stars are especially helium-rich, having lost most of
their hydrogen envelope through winds or interaction with a companion.
The range of progenitors that become WR stars is not well understood
and establishing a massive star evolution sequence represents one of the most serious challenges for modern stellar
theory \citep[see recent review by][]{crowther2007}.

Concrete mass and distance determinations are crucial to making
further progress, but this is difficult due to the large distances
from Earth at which most WR stars lie.  According to
\citet{vanderhucht2001}, there are only 19 WR stars with mass
estimates based on spectroscopic orbits and the vast majority of these
are for short-period systems with periods between 1--100 days.  Longer
period binaries are less likely to have had interactions between the
components but are difficult to characterize due to the lower orbital
speeds.

The subject of this Letter is WR~140 ($=$HD~193793), a WR binary
system with a 7.9~year period.  
\citet{williams1987} first noticed mysterious episodic infrared
variability and follow-up observations \citep{williams1990a,moffat1987}
established the cause to be dust creation near periastron of a highly
eccentric orbit, likely catalyzed in the colliding-wind interface
between the WR star and O-star winds \citep{usov91}.  Since this time,
WR~140 (WC7$+$O5) has been subject to many monitoring campaigns,
including infrared \citep{williams2009}, radio \citep{white1995}, and
in radial velocity \citep{fahed2011}.  Recently, progress towards a
proper visual orbit was made by the single-epoch detection of the
binary using the IOTA interferometer \citep{monnier2004} and through
repeated imaging of the rotating colliding-wind region using the VLBA
\citep{dougherty2005}.  Despite the wealth of data, mass estimates
have suffered from large errors ($\sim$20\%) due to lack of a
high-quality visual orbit to go along with precise spectroscopic data.

In this paper, we report seven epochs of binary observations at the near-infrared H and Ks bands with the IOTA and CHARA interferometers.  Our data span seven years allowing  us to construct the first complete visual orbit for WR~140. We combine this with recent spectroscopic work to determine precise masses and orbital parallax. 

\section{Observations}
\label{observations}

\subsection{IOTA}
Observations of WR~140 in 2003, 2004, and 2005 were obtained with the IOTA (Infrared-Optical
Telescope Array) interferometer \citep{traub2003}. IOTA 
was located on Mt. Hopkins (Arizona) and consisted of 
three 0.45-m telescopes that were movable among
17 stations along 2 orthogonal linear arms (telescopes A \& C could move
along the 35-m northeastern arm, while telescope B could move along the
15-m southeastern arm).  By observing a target in many different array
configurations, IOTA could synthesize an aperture 35m$\times$15m
(corresponding to an angular resolution of
$\frac{\lambda}{B}\sim$5$\times$12~milliarcseconds at 1.65$\mu$m).  
The observations of WR~140 from 2003 were first reported in
\citet{monnier2004}.

All of the IOTA observations included three simultaneous baselines
using the broad band H filter and the light beams from the three
telescopes were interfered using the single-mode IONIC3 combiner
\citep{berger2003}.  Basic data reduction procedures were the same as
described in several previous IOTA papers \citep[e.g.,][]{monnier2006} and
final error estimation followed the study of the $\lambda$~Vir binary
by \citet{zhao2007}; we applied 2\% relative and a $\Delta{\mathcal{V}^2}=0.02$ additive systematic
errors to all our measured visibility amplitude \viss data.  We found our closure phases
were partially corrupted by bandwidth smearing effects and we only
used \viss data for our orbit fitting \citep[see detailed description and
  simulation of these effects explored by][]{zhao2007}.
Table~\ref{table_obslog} contains a detailed log of the IOTA
observations.

\subsection{CHARA}
Observations of WR~140 in the years 2006, 2007, 2008, and 2009 were
obtained with the Center for High Angular Resolution Astronomy (CHARA)
Array.  CHARA is located on Mt. Wilson (California) and consists of
six fixed 1-m aperture telescopes with baselines ranging from 30 to
330m. CHARA is currently the longest baseline optical interferometer
in the world and can reach angular resolutions of $\sim$0.5
milli-arcseconds (mas) in the near-infrared.  The CHARA facility and
the two-beam CLASSIC combiner used in this work are described by
\citet{theo2005}.

Most of the CHARA observations were carried out in the Ks band
although a few data points were taken in the H band.  The observing
dates, wavelengths, and baselines can be found in
Table~\ref{table_obslog}.  We reduced the data using an IDL-based
suite of routines written by one of us (JDM) and these have been
previously described in \citet{tannirkulam2008}.  For this paper, we
adopted a 10\% relative and a $\Delta{\mathcal{V}^2}=0.02$ additive
error to account for systematic calibration errors\footnote{The calibration errors were lower for IOTA-IONIC3 because 
of the use of single-mode fibers while CHARA-CLASSIC is a ``free-space combiner'' with only limited spatial filtering.}.

We have reported all the calibrators and adopted sizes in Table~\ref{table_obslog}.
Uniform Disk (UD) diameters of interferometer calibrators were
generally estimated using {\em getCal}, an SED-fitting routine
maintained and distributed by the NASA Exoplanet Science Institute
(http://nexsci.caltech.edu), or SearchCal \citep{bonneau2006}, a surface
brightness algorithm maintained and distributed by the J. M.  Mariotti
Center (JMMC, http://jmmc.fr).  We note that {\em getCal} does not
produce as accurate diameter estimates as {\em searchCal} in general,
but {\em getCal} can be employed on a wider variety of calibrator
spectral types and was more useful in this current paper.

The final IOTA and CHARA calibrated \viss data were saved in OI-FITS format \citep{pauls2005} and
all files are available upon request.

\section{Analysis}
\label{analysis}
We have interpreted our \viss data using a binary model. Here we describe the components of the model and our fitting results.

\subsection{Description of binary model}
Our model for WR~140 consists of two stars, each modelled as a uniform
disk.  The size of the O-star can be estimated based on the effective
temperature and infrared flux estimate, while the WR-star size is
affected by the optically-thick wind \citep[we followed similar
  procedure as][to account for wind opacity]{millour2007}.  We found
that both stars are much smaller than the resolution of CHARA and so
our final results are not sensitive to our adopted UD size of 0.05~mas
for the WR star and 0.07~mas for the O-star.

The WR/O-star flux ratios at H band and Ks band were separately fitted
but held constant across all epochs.  The near-infrared flux
monitoring by \citet{taranova2011} showed that 52\% of the K-band
emission during the 2009 epoch was from the outburst dust shell and
was assumed to be over-resolved by CHARA.  No other epochs were 
affected by dust emission.  For a given set of orbital elements
($a$, $e$, $i$, $\omega$, $\Omega$, $P$, $T_0$), we can predict the separation and
position angle at the time of each observation.   The apparent brightness ratio can be
affected by the finite bandwidth of the observations (``bandwidth-smearing'') and
we have accounted for the square bandpasses of
the H and Ks filter using the basic procedure described in \citet{zhao2007}.  The
bandwidth-smearing correction was insignificant for the short baseline
IOTA data, but did affect the longer baseline CHARA data at the 5--10\% level.

For visualization purposes, we also collected each year's data and
fitted for relative positions.  The
best fitting locations of the O-star relative to the WR~star for each
observation year has been included in Figure~\ref{fig_orbit} and
allows comparison with our final orbit fitting results.  The size and
shape of each epoch's allowed region (an error ellipse containing the
68\% confidence region) vary significantly from year to year due to differences in the
quantity and quality of the \viss data. We also included the
parameters of these best-fit locations in Table~\ref{table_epochfits}.

To validate our choice of this simple binary model, we also carried
out model-independent image reconstructions for the 2003--2005
IOTA/IONIC3 data. We confirmed that the system is dominated by two
point sources.  We were interested to see if there could be any sign
of the colliding wind zone between the stars, but no extra emission
was seen in this region (at the level of a few percent of the peak
emission).  The brightest emission not coming from the two stars
showed up to the North-East of the system in some epochs at the 2--5\%
level -- likely an artifact from residual miscalibration.

\subsection{Orbit fitting procedure}
To arrive at our final orbital solutions, we fit directly to the \viss values.  We carried out two different fitting exercises that differed in how we incorporated spectroscopic data.
  
First we wanted to carry out an orbital fit as independent as possible
from the recent spectroscopic orbit of \citet{fahed2011}. This allows
us to independently confirm the crucial orbital elements $e$ and
$\omega$, although we adopted their values for the period ($P$) and
time of periastron ($T_0$) in this procedure.  The best-fitting
orbital elements (reduced $\chi^2=0.52$) are compiled in the right
column of Table~\ref{table_orbit}.  Error bars were estimated using
1000 bootstrap resamplings \citep{bootstrap} of the data (grouped by
night).  This is the same procedure recently applied for the visual
orbit of $\alpha$~Oph \citep{hinkley2011}. In order to capture the
uncertainties in the \citet{fahed2011} estimates for $P$ and $T_0$, we
did not strictly fix these quantities during the bootstrap fits but
rather used Monte Carlo sampling based on the \citet{fahed2011}
uncertainty estimates.

The  IOTA$+$CHARA visual orbit compares favorably with the spectroscopically-determined orbital elements from 
\citet{fahed2011}: 

\begin{itemize}
\item $e$: 0.8962$\pm$0.0014 (Fahed's orbit ``This paper+M03'') compared to eccentricity 0.901$^{+.006}_{-.004}$ (this work alone). These values are compatible and confirm the high orbit eccentricity.
\item $\omega$: 44.6\arcdeg$\pm$1.1\arcdeg (Fahed's orbit ``This orbit+M03'') compared to 48.2\arcdeg$\pm$1.3\arcdeg (this work alone). This is slightly discrepant, although note that one of the orbital solutions
presented in the \citet{fahed2011} report $\omega=47.5$\arcdeg (Fahed's orbit ``This paper''; middle solution in Table 2).
\end{itemize}

Much tighter constraints on the orbital elements can be attained by
using the spectroscopic values from \citet{fahed2011} as a {\em prior}
in the visual orbit fit (the orbit labeled ``This paper+M03'').  The most important effect of this is to
constrain the eccentricity, which is better determined by the
\citet{fahed2011} dataset that sampled the periastron period very
densely. Our visual observations missed the
fast-changing orbital motion in early 2009 and thus can not be expected to optimally
constrain eccentricity.

The first column of Table~\ref{table_orbit} contains our best-fitting
solution from the joint analysis (reduced $\chi^2=0.51$).
We used the same bootstrap procedure to determine the error bars on
each parameter.  To visualize the range of orbits allowed by our
solutions, we plotted all 1000 bootstrap orbits in
Figure~\ref{fig_orbit}.  As expected, the joint solution is better
constraining, especially near periastron.

The last step in our analysis was to take the new orbital elements and
use the spectroscopic $K$ values from \citet{fahed2011} to calculate
masses and orbital parallax. Strictly speaking, since the $K$ values
depend slightly on the orbital elements themselves, we re-fitted the
$\gamma$ and $K$ values for our orbital solutions using the radial
velocity data in \citet{fahed2011}. This refinement is slightly more
accurate than simply adopting the spectroscopic $a \sin i$ along with
astrometry to derive inclination and distance.  The final results for
masses and distance are also included in Table~\ref{table_orbit}.

Another way to view the remarkable quality of the fit is presented in
Figure~\ref{fig_datafit}. Here we show the observed \viss as a
function of the projected separation of the two components (joint
orbital solution).  This figure also shows the effect of bandwidth
smearing for large projected separations.

\section{Discussion and Future Work}
\label{discussion}

The WR and O-star masses found here are similar to previous estimates
\citep[e.g.][]{vanderhucht2001} but with about 6$\times$ smaller error bars
($\sim$3--4\%). As has been seen in  other WC binary systems,  we find the
WR star mass is less than 20$\msun$
and less than half the current mass of the O-star companion.  These
two qualities fit the interesting trend seen in the (six) WC stars
with mass estimates \citep{crowther2007}.  Our accurate distance will
help to place these two stars in the Hertzsprung-Russel diagram for a
stringent test of stellar evolution models that include
mass-loss. This is only the second galactic WR star to have a distance
measured through (orbital) parallax \citep[the other is
  $\gamma$~Vel;][]{millour2007}.  Note that the periastron distance of
the orbit is 1.53~AU, which means the O-star companion is too distant
to have affected the stellar evolution of the Wolf-Rayet star --
unless the WR star progenitor experienced a red supergiant stage
\citep[see case of WR~104;][]{tuthill2008} or the orbit has
drastically changed due to mass-loss.  

The interferometry data allow us to measure the 
flux ratios of the two stars ($\frac{ {\rm  Flux}_{\rm (WR)}}{ {\rm Flux}_{\rm (O)}}$) 
in the H- and Ks-bands for  the first time.
Our H-band IOTA data employed closure phases, allowing us to identify 
the Northwest component to be brightest in 2003.  We
have had to assume that this component is also brighter at K band
since we lack closure phase data to break the 180\arcdeg~ degeneracy of
single-baseline data.  

Based on our flux ratios, the IR-bright component has a significantly 
redder color than the IR-faint component, consistent with the IR-bright component being the WR star.
The shape of the non-thermal radio
emission \citep{dougherty2005} also identifies the Northwest
(IR-bright) component in 2003 to be the WR star that is expected to possess the
the higher-momentum wind.

Armed with this knowledge, we can decompose the combined spectral
energy distribution (SED) into their component SEDs.  For the combined
system, we adopt (non-dusty) BVJHK magnitudes $ = (7.28, 6.89, 5.71,
5.35, 5.02)$ \citep{reed2003,taranova2011} and assume the O5III star
has colors given by \citet{martins2006}. The last ingredient we need
is the least certain: the interstellar reddening.  We will adopt the
reddening law of \citet{mathis1990} with $R_V$=3.1 and with $A_V$
spanning $A_V=2.95$ from \citet{morris1993} using the 2175\AA\
feature and $A_V=2.06$ from SED colors
\citep{conti1990}\footnote{Here, we have used the relation that
  $A_v=1.1~A_V$ \citep[e.g.,][]{smith1968}} -- as you will see, the
large uncertainty in $A_V$ leads to large errors in our system
luminosity despite our new well-constrained distance.  We can now fix
the H band ratio to 1.37 from our IOTA data and solve for the
remaining flux ratios (modulo the $A_V$ uncertainty).  For $A_V=2.95$,
the flux ratios at BVJHK become $(2.46, 1.64, 1.05, 1.37, 1.93)$; for
$A_V=2.06$ the flux ratios become $(0.37, 0.35, 0.88, 1.37, 2.09)$.
The observed flux ratio at Ks band (1.94$\pm$0.06) is within the range seen here
and is in good agreement with the high $A_V$ case.
Depending on which lines are used and which templates are adopted,
\citet{fahed2011} argue that the O-star might range from $\sim$0.5 to $\sim$3$\times$ brighter than the WR star in
the optical band continuum -- a wide range compatible with our derived flux ratios.
Our prediction for V band flux rato varies by a factor of
5 (!) because of $A_V$ uncertainties and thus an interferometric
measurement here could be exploited to strongly constrain the true
reddening, a crucial component to luminosity determination which wei
now explore.  

Using our new distance ($d=1.67$~kpc) and the range of flux ratios, we
find $M_V$ for the O-star ranges from -6.11 and -5.94 and the $M_V$ for
the WR star ranges from -6.6 and -4.8.  This would classify the O5
star as intermediate between giant and supergiant
\citep[cf.][]{martins2006}. The WR star luminosity class is hardly
constrained but the high $A_V$ case yields unrealisticly high
luminosities as judged by other WC7 stars with known distances
\citep[via cluster membership;][]{vanderhucht2001}. 
In an attempt to reconcile the differences, we scaled the fluxes 
for a 70kK CMFGEN model for the WC7 star from \citet{smith2002} to match the measured flux ratio of 1.37 in 
the H Band using a model SED for the O5 star from \citet{martins2005}.
The WR/O5 flux ratios in $V$ and $v$ were found to be 1.05 and 0.73 
respectively, that in $V$ being affected by strong emission lines.
The absolute magnitudes derived for the O5 star, $M_V$ = -6.37 and 
-5.65 for high and low reddening extremes, are near or above those 
for supergiants, whereas those for the WC star, $M_v$ = -6.7 and 
-5.9 are both anomalously high. 
Resolution of the uncertain interstellar reddening of WR~140 is 
beyond the scope of the present paper but is urgently required to 
exploit the determination of its parallax.

We expect our success here will motivate future observations of
galactic Wolf-Rayet stars using today's interferometers from visible
to NIR wavelengths.  The multi-wavelength flux ratios can help yield
crucially-needed new estimates of interstellar reddening and the
possibility to combine spectroscopic orbits with new interferometric
visual orbits will allow accurate distance, mass, and luminosity
measurements for a substantially larger set of galactic WR stars.
Such a dataset is important to test the current massive stellar
evolution paradigm that tells us how main-sequence O-stars move
through the various stages of Red Supergiant, Wolf-Rayet (WN \& WC)
and Luminous Blue Variable before ultimately becoming a Supernova.

\acknowledgments { We have appreciated discussions with Tony Moffat,
  Peter Tuthill, Debra Wallace, Bill Danchi, Sean Dougherty, and Remi
  Fahed during the (long) course of this work.  We thank SAO, U. Mass,
  NSF AST-0138303, NSF AST-0352723, and NASA NNG05G1180G for
  supporting IOTA development and operations.  We also acknowledge
  funding from GSU, the Keck Foundation, and NSF AST-0908253 for the
  CHARA Array.  IONIC-3 was developed by LAOG (now IPAG) and LETI in
  the context of the IONIC collaboration (LAOG, IMEP, LETI), funded by
  the CNRS and CNES (France).  Lastly we thank NSF
  AST-0807577 for support of University of Michigan researchers in
  this work.  EP received funding from a Michelson Postdoctoral
  Fellowship and a Scottish Universities Physics Alliance (SUPA)
  advanced fellowship.  PMW is grateful to the Institute for Astronomy
  for hospitality and continued access to the facilities of the Royal
  Observatory, Edinburgh.  This research has made use of the SIMBAD
  database, operated at CDS, Strasbourg, France, and NASA's
  Astrophysics Data System (ADS) Bibliographic Services.

{\it Facility:} \facility{IOTA (IONIC3), \facility{CHARA (CLASSIC)}
}

\bibliographystyle{apj}

\clearpage
\begin{deluxetable}{cllcc}
\tabletypesize{\scriptsize}
\tablecaption{Observing log for WR140. All calibrated OI-FITS data available upon request.
\label{table_obslog}}
\tablewidth{0pt}
\tablehead{ \colhead{Orbital} & \colhead{Date} &  \colhead{Interferometer} & \colhead{$\lambda_0$} & \colhead{Bandwidth} \\ 
\colhead{Phase\tablenotemark{a}} & \colhead{(UT)} & \colhead{(Configuration)} &\colhead{($\mu$m)} &  \colhead{($\mu$m)} }
\startdata
 2.296 & 2003Jun17 & IOTA\tablenotemark{b} (A35C15C10) & 1.650 & 0.248 \\ 
2.406 & 2004Apr30 & IOTA (A35B15C10) & 1.650 & 0.248 \\ 
2.407 & 2004May01 & IOTA (A35B15C10) & 1.650 & 0.248 \\
 2.417 & 2004May30 & IOTA (A35B15C10) & 1.650 & 0.248 \\
 2.417 & 2004Jun01 & IOTA (A35B15C10) & 1.650 & 0.248 \\ 
2.418 & 2004Jun04 & IOTA (A35B15C10) & 1.650 & 0.248 \\ 
2.419 & 2004Jun05 & IOTA (A35B15C10) & 1.650 & 0.248 \\
 2.419 & 2004Jun06 & IOTA (A35B15C10) & 1.650 & 0.248 \\ 
2.547 & 2005Jun11 & IOTA (A35B15C10) & 1.650 & 0.248 \\ 
2.547 & 2005Jun13 & IOTA (A35B15C10) & 1.650 & 0.248 \\
 2.548 & 2005Jun14 & IOTA (A35B15C10) & 1.650 & 0.248 \\
 2.548 & 2005Jun15 & IOTA (A35B15C10) & 1.650 & 0.248 \\
 2.548 & 2005Jun16 & IOTA (A35B15C10) & 1.650 & 0.248 \\
 2.549 &
2005Jun17 & IOTA (A35B15C10) & 1.650 & 0.248 \\
 2.549 & 2005Jun18 &
IOTA (A35B15C10) & 1.650 & 0.248 \\
 2.698 & 2006Aug22 &
CHARA\tablenotemark{c} (W2-E2) & 2.133 & 0.350 \\
 2.799 & 2007Jun11
& CHARA (W2-E2) & 1.673 & 0.274 \\
 2.800 & 2007Jun13 & CHARA (W2-E2) &
2.133 & 0.350 \\
 2.926 & 2008Jun14 & CHARA (W2-E2) & 1.673 & 0.274 \\
 2.927 & 2008Jun16 & CHARA (W2-S1) & 2.133 & 0.350 \\
 2.927 &
2008Jun17 & CHARA (W1-S2) & 2.133 & 0.350 \\ 
2.928 & 2008Jun18 & CHARA
(W1-S2) & 2.133 & 0.350 \\ 3.054 & 2009Jun20 & CHARA (W1-S2) & 2.133 &
0.350 \\ 3.055 & 2009Jun22 & CHARA (W1-S2) & 2.133 & 0.350 \\ 3.055 &
2009Jun23 & CHARA (S2-E2) & 2.133 & 0.350 \\ 3.056 & 2009Jun25 & CHARA
(S2-E2) & 2.133 & 0.350 
\enddata   
\tablenotetext{a}{Orbital phase assuming $T_0 = 2446156.2$
  (MJD), $P = 2896.5$ days \citep{fahed2011}.}  \tablenotetext{b}{IOTA
  calibration employed the following calibrators (all sizes were
  estimated using getCal): HD~192985 (0.46$\pm$0.06 mas), HD 193631
  (0.31$\pm$0.28 mas), HD~126035 (0.78$\pm$0.24 mas), HD~193664
  (0.58$\pm$0.05 mas), HD~193961 (0.24$\pm$0.06 mas) }
\tablenotetext{c}{CHARA calibration employed the following calibrators
  (all sizes estimated using getCal, except HD196360): HD~192985
  (0.46$\pm$0.06 mas), HD~193631 (0.31$\pm$0.28 mas), HD~196360
  (0.61$\pm$0.05 mas), HD~192640 (0.40$\pm$0.15 mas), HD~195194
  (0.63$\pm$0.13 mas)}
\end{deluxetable}

\clearpage

\clearpage
\begin{deluxetable}{l|ll|lll}
\tabletypesize{\scriptsize}
\tablecaption{Position of O-star with respect to WR star
\label{table_epochfits}}
\tablewidth{0pt}
\tablehead{
\colhead{Mean Date}  & \multicolumn{2}{c}{{Relative Position}}  & \multicolumn{3}{c}{{Error Ellipse\tablenotemark{a}}} \\
\colhead{(UT)} & \colhead{East (mas) } & \colhead{North (mas)} & \colhead{Major (mas)} &
\colhead{Minor (mas)} & \colhead{PA Major (E of N)}
} 
\startdata
2003Jun17&  5.70& -11.31&  0.87&  0.30&  -86\\
2004May25&  6.87& -11.33&  0.34&  0.11&  -61\\
2005Jun15&  7.36& -10.16&  0.37&  0.11&  -57\\
2006Aug22&  6.97&  -7.14&  4.85&  0.26&  112\\
2007Jun12&  5.73&  -4.18&  1.25&  0.10&  -28\\
2008Jun17&  3.51&  -0.76&  0.11&  0.05&   51\\
2009Jun23&  0.86&  -5.46&  0.70&  0.13&  -47
\enddata
\tablenotetext{a}{Error ellipse contains 68\% confidence interval and is specified by the $\pm$error in the two orthogonal directions (in milliarcseconds) specified by the position angle of the ellipse major axis (degrees East of North). See Figure~\ref{fig_orbit} for graphical representation.}
\end{deluxetable}

\clearpage
\begin{deluxetable}{lcc}
\tablecaption{Orbital parameters for WR140
(adopting WR as primary star, O as secondary star)
\label{table_orbit}
}
\tablewidth{0pt}
\tablehead{
\colhead{Parameter} & \colhead{This work with} & \colhead{This work alone\tablenotemark{a}} \\
 &  \colhead{Fahed et al. (2011) prior} &
}
\startdata
Flux Ratio\tablenotemark{b} (H band) & \multicolumn{2}{c}{1.37$\pm$0.03} \\
Flux Ratio (Ks band)                  & \multicolumn{2}{c}{1.94$\pm$0.06} \\
Semi-Major Axis (mas)                & 8.82$\pm$0.05 & 8.99$^{+0.16}_{-0.22}$ \\
Eccentricity                         & 0.8964$^{+0.0004}_{-0.0007}$ & 0.901$^{+.006}_{-.004}$ \\
Inclination (deg)                    & 119.6$\pm$0.5 & 118.9$^{+1.3}_{-0.5}$ \\
$\omega$ (deg)                       & 46.8$\pm$0.4  & 48.2$\pm$1.9 \\
$\Omega$ (deg)                       & 353.6$\pm$0.4 & 354.2$^{+0.9}_{-0.5}$ \\
Period (days)                        & 2896.35$\pm$0.20 & (2896.5$^{+0.2}_{-1.5}$)\\
T$_0$ (MJD)                          & 46154.8$\pm$0.8 & (46155.7$^{+2.6}_{-3.3}$)\\
$\sum{\chi^2}$/DOF		     & 0.51 & 0.52 \\
\hline
\multicolumn{3}{l}{Derived Physical Quantities\tablenotemark{c}} \\
\hline
Distance (kpc)			     & 1.67$\pm$0.03 & 1.60$^{+0.11}_{-0.07}$ \\
M$_{\rm WR}$ (\msun)                   & 14.9$\pm$0.5 & 13.9$^{+1.9}_{-1.2}$ \\
M$_{\rm O}$ (\msun)                    & 35.9$\pm$1.3 & 33.1$^{+4.5}_{-2.8}$ 
\enddata
\tablenotetext{a}{All orbital elements were fitted to the visual orbit date presented here, except 
for the $P$ and $T_0$, which were adopted from \citet{fahed2011}.}
\tablenotetext{b}{Flux ratio is $\frac{\rm WR~ flux~ density}{\rm O5~ flux~ density}$.}
\tablenotetext{c}{The visual orbit above and the spectroscopic data from \citet{fahed2011}
were combined to derive the orbital parallax and relative masses. We emphasize that the
high precision on the mass and distance requires both the visual orbit presented here and 
high quality spectroscopic data of \citet{fahed2011}.}
\end{deluxetable}

\clearpage

\begin{figure}[hbt]
\begin{center}
\includegraphics[angle=0,width=6in]{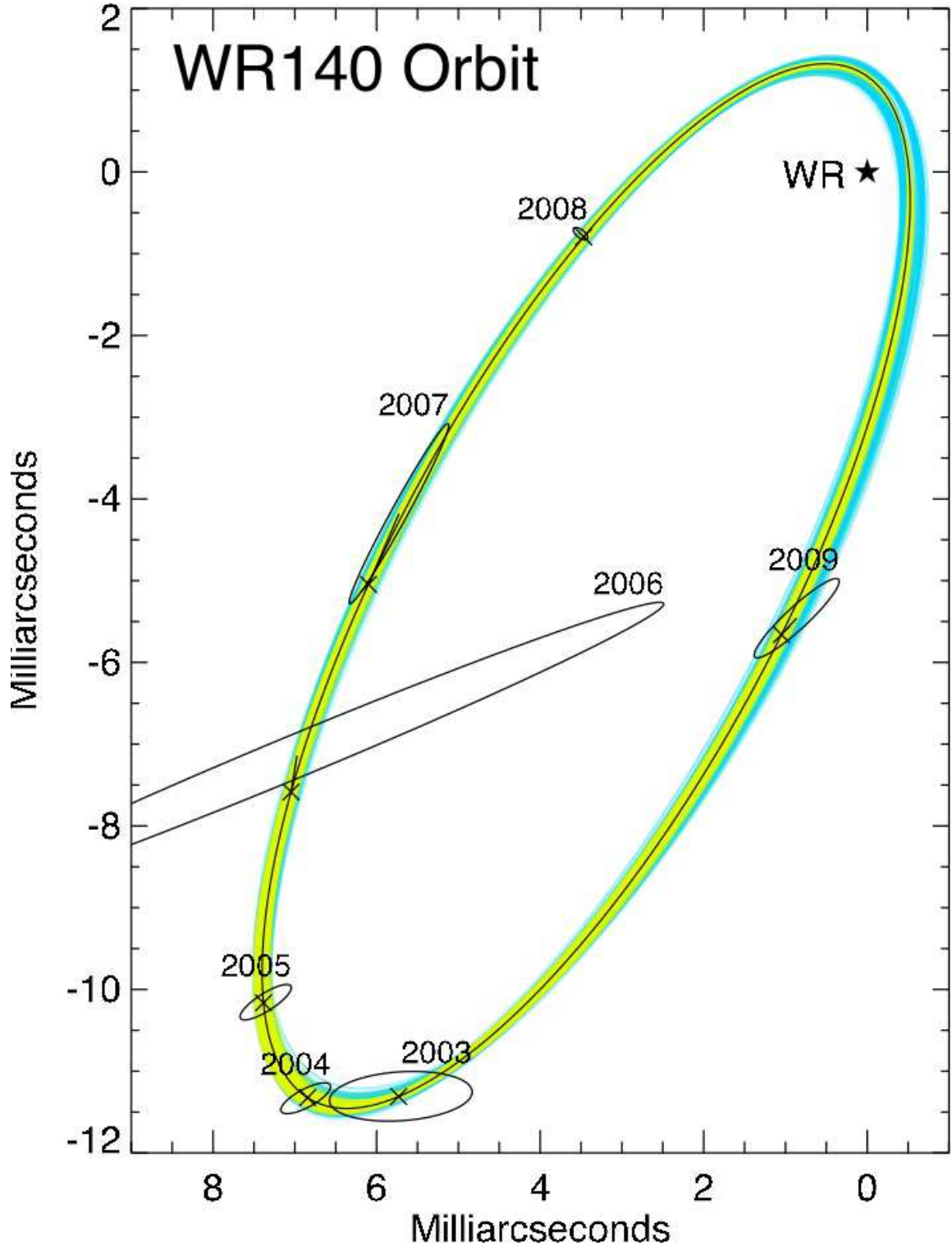}
\figcaption{\footnotesize  This figure shows the 3-$\sigma$ band of allowed orbits based on our visual data alone (thick blue band) and for the more-constraining, joint visual/spectroscopic solution (yellow band).  The best fit joint solution is shown with solid line.  For each year, the data were analyzed separately and the positions of the O star with respect to WR star are shown here, marked by error ellipses (see Table~\ref{table_epochfits}). The best-fit orbit prediction for each epoch is connected to each correponding error ellipse.
\label{fig_orbit}}
\end{center}
\end{figure}

\begin{figure}[hbt]
\begin{center}
\includegraphics[angle=90,width=6in]{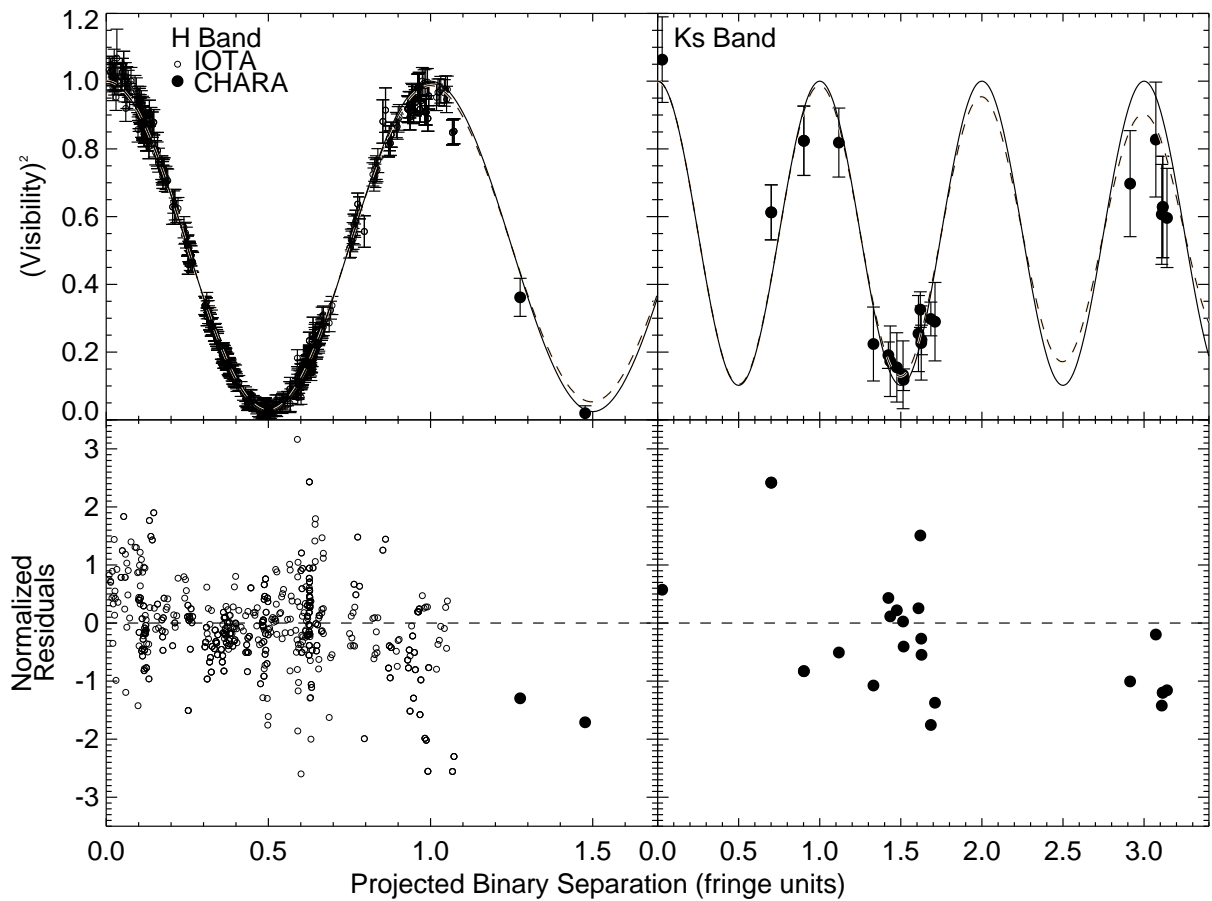}
\figcaption{\footnotesize 
Here we plot the observed visibility data as a function of the projected binary separation of our best-fitting joint model. We have plotted the expected curve both for no bandwidth smearing (solid line) and for
the appropriate level we used in this work (dashed line).The bottom panels shows the residuals between data and model normalized by the data errors.  Note the visibility for the CHARA 2009 data was boosted by factor of 2.1 to account for the extra
over-resolved emission by the short-lived dust shell created at periastron \citep[48\% stellar emission, 52\% dust;][]{taranova2011}.
\label{fig_datafit}}
\end{center}
\end{figure}

\end{document}